\documentclass[onecolumn, prl, aps, superscriptaddress, showpacs]{revtex4}
\usepackage{amsmath}
\usepackage{amssymb}
\usepackage{graphicx}
\usepackage{bbm}
\usepackage{color}

\begin{document}

\title{Dynamical mean field solution of the Bose-Hubbard model}
\author{Peter Anders}
\affiliation{Theoretische Physik, ETH Zurich, 8093 Zurich, Switzerland}
\author{Emanuel Gull}
\affiliation{Department of Physics, Columbia University, 538 West 120th Street, New York, NY 10027, USA}
\author{Lode Pollet}
\affiliation{Department of Physics, Harvard University, Cambridge, Massachusetts 02138, USA}
\author{Matthias Troyer}
\affiliation{Theoretische Physik, ETH Zurich, 8093 Zurich, Switzerland}
\author{Philipp Werner}
\affiliation{Theoretische Physik, ETH Zurich, 8093 Zurich, Switzerland}
\date{\today}

\begin{abstract}
We present the effective action and self-consistency equations for the bosonic dynamical mean field (B-DMFT) approximation to the bosonic Hubbard model and show that it provides remarkably accurate phase diagrams and correlation functions. To solve the bosonic dynamical mean field equations we use a continuous-time Monte Carlo method for bosonic impurity models based on a diagrammatic expansion in the hybridization and condensate coupling. This method is readily generalized to bosonic mixtures, spinful bosons, and Bose-Fermi mixtures.
\end{abstract}



\maketitle

Dynamical mean field theory (DMFT) is a computationally tractable framework for the study of fermionic lattice models, which becomes exact in the limit of infinite dimensions or infinite coordination number \cite{Metzner89, Georges92, Georges96}. In this limit the self-energy is momentum independent and can be obtained from the solution of an appropriately defined impurity problem. In finite dimensions, the approximation of a momentum-independent self-energy means that spatial correlations are neglected, but the local dynamics can be fully taken into account. DMFT provides a powerful, non-perturbative tool to investigate correlation effects and has been used extensively to study the properties of strongly correlated electron systems \cite{Georges96, Kotliar06}. The DMFT formalism is particularly well-suited to study the Mott transition, which is driven by local physics. 

The formulation of a dynamical mean field theory for bosonic lattice models is related to the formulation of an extended-DMFT for the ordered phase \cite{Chitra01}. Attempts to derive bosonic DMFT equations based on the infinite coordination limit face the problem that normal and condensed bosons must be scaled differently, rendering the scaling ambiguous (in fact the pathological infinite coordination limit can only be defined for a classical field~\cite{Fisher89}). In a recent paper \cite{Byczuk08} Byczuk and Vollhardt suggested to perform the scaling in the action rather than in the Hamiltonian. They obtained a conventional DMFT description with a momentum independent self-energy which requires the self-consistent solution of an impurity problem which couples to two baths, a bath of normal bosons and a bath of condensed bosons. While Ref.~\onlinecite{Byczuk08} presented results for the bosonic Falicov-Kimball model, other groups applied (variants of) the B-DMFT formalism to the Bose-Hubbard model \cite{Hu09} and multi-component Bose gases \cite{Hubener09}.
However, these previous formulations of B-DMFT were either incorrect or incomplete. 
 
Here we derive internally consistent B-DMFT equations for the Bose-Hubbard model, which produce physically meaningful solutions over the whole parameter range, and recover the limits of the dilute Bose gas~\cite{CapogrossoSansone09}  and non-interacting bosons. By comparing to the numerically exact lattice QMC simulation~\cite{Prokofev98} of the full lattice model we show that our formalism yields remarkably accurate results~\cite{MC} both for phase diagrams and correlation functions. The DMFT equations for fermionic models can be derived (without any reference to an infinite coordination limit) using functionals of the local Green's function \cite{Antoine}, with  DMFT corresponding to a certain approximation of the kinetic energy functional. Our derivation of the B-DMFT equations is completely analogous~\cite{EPAPS}.   
 
In both the fermionic and bosonic versions of DMFT, the computationally challenging part is the solution of a quantum impurity problem. For fermionic impurity models, significant progress has been made with the development of continuous-time Monte Carlo techniques, based on an expansion of the partition function in powers of the interaction \cite{Rubtsov05, Gull08} or the impurity-bath hybridization \cite{Werner06}. 
In this Letter we show that a similar hybridization-expansion approach yields an efficient quantum Monte Carlo algorithm for bosonic impurity models. 

Our specific model is that of spinless bosons on a three-dimensional (3D) simple cubic lattice with Hamiltonian
\begin{equation}
H = - t \sum_{\langle i,j\rangle} b_i^{\dagger}b_j +\frac{U}{2}\sum_i n_i (n_i - 1) - \mu \sum_i n_i,
\label{hamiltonian}
\end{equation}
where $t$ denotes the hopping amplitude, $U$ the on-site interaction and $\mu$ the chemical potential. 

The effective impurity model of B-DMFT contains the microscopic local terms of the Hamiltonian ($U$ and $\mu$) to which two additional source fields are added. The first one can, like in static mean-field theory, be written as $-zt \phi$ (with $\phi$ the constant condensate) and is conjugate to the $b$ and $b^{\dagger}$ operators, such that $\langle b \rangle$ can become a non-zero complex number.  Fluctuations around the condensate at the one-loop level can be added after decomposing $b(\tau) = \langle b \rangle + \delta b(\tau)$. The source field for the two-particle channel couples to {\it non-condensed} operators~\cite{NegeleOrland} of the form $\delta b^{\dagger}(\tau)  \delta b (\tau') $ (and also  $\delta b(\tau)  \delta b (\tau') $ because of the symmetry breaking). This follows from a functional derivation~\cite{EPAPS} in which the {\it connected} local Green's function is constrained \cite{footnote}. If the full local Green's function were constrained, one would not arrive at the correct B-DMFT equations, because the approximation of the kinetic energy functional would mix up small and large contributions. 

Writing the partition function as $Z={\rm Tr}_b[Te^{-S_\text{imp}}]$ and shifting contributions between the hybridization and condensate terms, our final action can be expressed using the full operators $b$ in the (Nambu) form:

\begin{align}
&S_\text{imp}=- \frac{1}{2}\int_0^{\beta} d\tau d\tau' \textbf{b}^{\dagger}(\tau) \mathbf{\Delta}(\tau-\tau') \textbf{b}(\tau')- \tilde \mu \int_0^{\beta} d\tau n(\tau)\nonumber\\
&+\frac{U}{2} \int_0^{\beta} d\tau n(\tau)[n(\tau)-1]- \kappa \mathbf{\Phi}^{\dagger} \int_0^{\beta} d\tau  \textbf{b}(\tau).
\label{action}
\end{align}
The boson creation/annihilation operators are $\mathbf{b}^\dagger=(b^\dagger, b)$, the time-independent condensate is $\mathbf{\Phi}^\dagger=(\phi^*, \phi)$, and the hybridization function $\mathbf{\Delta}$ is related to the mean-field propagator $\mathbf{G_0}$ through
\begin{equation}
\mathbf{\Delta}(i \omega_n) = -i \omega_n \sigma_3 - \tilde{\mu} \mathbf{1} + \mathbf{G}_{0}^{-1}(i \omega_n).
\label{hyb}
\end{equation}
The parameter $\tilde{\mu} = \mu - \langle\epsilon\rangle$ is chosen such that
$\mathbf{\Delta}(i \omega_n) \rightarrow 0$ in the limit $\omega_n
\rightarrow \infty$. We will consider here a symmetric density of states, where $\langle \epsilon \rangle = 0$, and just write $\mu$ from now on. We furthermore define the elements of the hybridization matrix as
\begin{eqnarray}
\mathbf{\Delta}(\tau-\tau')= \left( \begin{array}{cc}
F(\tau'-\tau)  & 2K(\tau-\tau') \\
2K^{*}(\tau-\tau') & F(\tau-\tau')
\end{array} \right).
\end{eqnarray}
The condensate $\mathbf {\Phi}$  is constant in time and determined by the simple self-consistency condition
\begin{equation}
\label{eq:phi}
\mathbf{\Phi} = \langle \mathbf{b}(\tau) \rangle_{S_{\text{imp}}}.
\end{equation}
To determine the hybridization function $\mathbf{\Delta}$ we calculate the matrix self energy using the Dyson equation
\begin{equation}
\mathbf{\Sigma}(i\omega_n) = \mathbf{G}_0^{-1}(i\omega_n) - \mathbf{G}_{c}^{-1}(i\omega_n),
\end{equation}
where the connected part of the Green's function for the normal bosons is given by 
$\mathbf{G}_{c}(\tau) =-\langle T \textbf{b}(\tau) \textbf{b}^{\dagger}(0)  \rangle_{S_{\text{imp}}} + \mathbf{\Phi}\mathbf{\Phi}^{\dagger}$.
From $\mathbf{\Sigma}$ and the dispersion $\epsilon_{\mathbf{k}}$ of the lattice we obtain the local lattice Green's function
\begin{equation}
\mathbf{G}_\text{latt}(i\omega_n) = \sum_{\mathbf{k}}\Big[i\omega_n\sigma_3+(\mu-\epsilon_{\mathbf{k}})\mathbf{1} -\mathbf{\Sigma}(i\omega_n)\Big]^{-1},
\end{equation}
and the self-consistency condition requires that the impurity Green's function coincides with the local lattice Green's function: 
\begin{equation}
\label{eq:self-consistency}
\mathbf{G}_{0}^{-1}(i\omega_n) = \mathbf{\Sigma}(i\omega_n)  + \mathbf{G}_\text{latt}^{-1}(i\omega_n).
\end{equation}

Assuming $K=K^*$ and $\phi=\phi^*$, the coupling  $\kappa$ between the impurity and the condensate is given by \cite{EPAPS}
\begin{equation}
\label{eq:kappa}
\kappa = zt - F(i\omega_n=0) -2 K(i\omega_n=0).
\end{equation}
If one incorrectly uses $\kappa=zt$~\cite{Byczuk08} the trivial limit of the dilute Bose gas~\cite{CapogrossoSansone09} and the non-interacting Bose gas is incorrectly reproduced in finite-dimensional systems, while the condensed phase becomes unstable in a wide region of parameter space.
Hu and Tong \cite{Hu09} set the off-diagonal hybridization function $K$ to zero after each iteration to avoid these instabilities, and thus solved incorrect  B-DMFT equations, which equally fail to reproduce  the trivial limits. In Ref. \cite{Hubener09}, B-DMFT was  considered as an expansion in $1/z$ on a tree-like structure and a perturbative prescription was used to rescale the condensate (or $\kappa$) after each step. Complementing their $1/z$ expansion approach with the correct non-perturbative prescription would lead to Eqs.~(\ref{action}) and (\ref{eq:kappa})~\cite{Semerjian09}.
 
Our action reproduces the correct results in all limits on any lattice, and also produces stable B-DMFT solutions including in phases with a non-zero condensate. In the non-interacting model,  the chemical potential is pinned at the lower band edge in the presence of a finite condensate and Eq.~(\ref{eq:kappa}) reduces to $\kappa=-G_0^{-1}(i\omega_n=0)$. Note that there is an isolated state at energy $-zt$ for the Bethe lattice DOS \cite{Semerjian10}. In the static case without hybridization Eq.~(\ref{eq:kappa}) gives $\kappa=zt$ consistent with static mean field theory.  In the limit of infinite dimensions, only static mean-field theory is physical, in which case the addition of the source field $\mathbf{\Delta}$ is impossible because $\delta b = 0$.
In fact, the entire phase diagram of the 3D Bose-Hubbard model, as well as dynamical quantities such as correlation functions, are reproduced by the B-DMFT action with remarkable accuracy.

The self-consistency equations are solved by starting from an initial guess, solving the quantum impurity problem and then calculating new values for $\mathbf{\Phi}$ through Eq.~(\ref{eq:phi}), $\mathbf{\Delta}$ through Eq.~(\ref{eq:self-consistency}) and $\kappa$ through Eq.~(\ref{eq:kappa}). This procedure is repeated until convergence is reached. 

\begin{figure}[t]
\begin{center}
\includegraphics[angle=0, width=0.5\columnwidth]{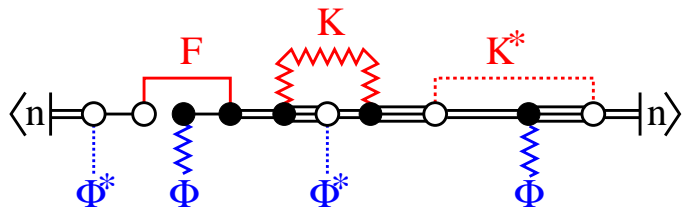}
\caption{Diagram corresponding to perturbation orders $m_F=1$, $m_K=1$, $m_{K^*}=1$, $m_\phi=2$, $m_{\phi^*}=2$ and $n(\tau=0)=2$. 
}
\label{diagrams}
\end{center}
\end{figure}

The computationally demanding step
is the solution of the bosonic quantum impurity problem and we will now present a quantum Monte Carlo (QMC) algorithm for its solution, which is similar in spirit to the fermionic hybridization expansion algorithm of Ref.~\cite{Werner06}.  
We expand the partition function $Z=\text{Tr}_b [T e^{-S_\text{imp}}]$ in powers of the hybridization functions $F$, $K$, $K^*$ and the source fields $\phi$ and $\phi^*$. 
This leads to an expression for the partition function as a sum of diagrams of the type illustrated in Fig.~\ref{diagrams}, which can be represented by a collection of $m_F+2m_{K^*}+m_\phi$ creation operators and the same number $m_F+2m_K+m_{\phi^*}$ of annihilation operators on the imaginary time interval $[0,\beta)$. Hybridization functions $F$ connect $m_F$ pairs of creation and annihilation operators, off-diagonal hybridization function $K$ ($K^*$) connect $m_K$ ($m_{K^*}$) pairs of creation (annihilation) operators, while $m_\phi$ ($m_{\phi^*}$) creation (annihilation) operators are linked to source fields $\phi$ ($\phi^*$). 
The integer $n\ge 0$ corresponds to the occupation of the impurity at times $\tau=0$ and $\beta$, and thus fixes $n(\tau)$.

An ergodic sampling of all possible diagrams requires the following updates:
(i) insertion/removal of a pair $b(\tau)F(\tau-\tau')b^\dagger(\tau')$,
(ii) increase/decrease of $n$ by one,
(iii) change of the bath type: 
\begin{align}
b(\tau)F(\tau-\tau')b^\dagger(\tau') &\leftrightarrow \kappa\phi^*b(\tau)\kappa\phi b^\dagger(\tau'),\\
b(\tau)K^*(\tau-\tau')b(\tau') &\leftrightarrow \kappa\phi^* b(\tau)\kappa\phi^* b(\tau'),\\
b^\dagger(\tau)K(\tau-\tau')b^\dagger(\tau') &\leftrightarrow \kappa\phi b^\dagger(\tau)\kappa\phi b^\dagger(\tau').
\end{align}
%
Additional updates such as shifts of operator times and reconnections of hybridization lines can be used to improve the efficiency. Denoting the trace contribution of a diagram $\langle n| \ldots | n\rangle$ by $w_{Tr}(n;\tau^F_1,\ldots, \tau^F_{m_F},\tau'^F_1,\ldots, \tau'^F_{m_F}; \ldots)$, the detailed balance condition for inserting/removing a pair $b(\tau)F(\tau-\tau')b^\dagger(\tau')$ becomes
\begin{align}
&\frac{p(m_F\rightarrow m_F+1)}{p(m_F+1 \rightarrow m_F)}=\frac{\beta^2}{m_F+1}F(\tau-\tau')\nonumber\\
&\hspace{5mm}\times \frac{w_{Tr}(n;\tau^F_1,\ldots,\tau,\ldots, \tau^F_{m_F},\tau'^F_1,\ldots, \tau', \ldots \tau'^F_{m_F};\ldots)}
{w_{Tr}(n;\tau^F_1,\ldots, \tau^F_{m_F},\tau'^F_1,\ldots, \tau'^F_{m_F};\ldots)},
\end{align}
if we choose the times $\tau$ and $\tau'$ randomly in $[0,\beta)$ and propose to remove this pair with probability $1/(m_F+1)$. Similar formulas can be derived for the other updates.

By taking functional derivatives of the partition function with respect to either $F(\tau-\tau')$, $K(\tau-\tau')$, $K^{*}(\tau-\tau')$, or $\phi^{*}$ and $\phi$ one obtains measurement formulas for the diagonal and off-diagonal parts of the Green function matrix and the condensate order parameter:
\begin{eqnarray}
\langle b(\tau)b^\dagger(0)\rangle_{S_\text{imp}} &=& \Bigg\langle \sum_{i=1}^{m_F} \frac{\Delta(\tau,\tau_i^F - \tau_i'^F)}{\beta F(\tau_i^F - \tau_i'^F)} \Bigg\rangle_{MC},\\
\langle b(\tau)b(0)\rangle_{S_\text{imp}} &=& \Bigg\langle \sum_{i=1}^{m_{K^*}}
\frac{\Delta(\tau, \tau_i^{K^*} - \tau_i'^{K^*})}{\beta K^*(\tau_i^{K^*} -
\tau_i'^{K^*})}\Bigg\rangle_{MC},\hspace{5mm}\\
\langle b(\tau) \rangle_{S_\text{imp}} &=& \Bigg\langle \sum_{i=1}^{m_{\phi^*}} \frac{\Delta(\tau,
\tau_i^{\phi^*})}{\kappa \phi^{*}} \Bigg\rangle_{MC}, 
\end{eqnarray}
and similarly for the adjoint with $\Delta(\tau, \tilde{\tau}) = \delta(\tau-\tilde{\tau})$ for $\tilde{\tau}\geq 0$ and $\delta(\tau-\tilde{\tau}-\beta)$ for $\tilde\tau < 0$.

The end point of $G(\tau)=-\langle Tb(\tau)b^\dagger(0)\rangle_{S_\text{imp}}$ is given by the density
$G(\beta_{-})=-\left\langle n \right\rangle_{MC} = -\langle \frac{1}{\beta} \int_0^{\beta} d\tau n(\tau)\rangle_{MC}$
and $G(0_{+})=G(\beta_{-}) - 1$.
Here $\langle A\rangle_{MC}$ means that the quantity $A$ should be averaged over all configurations obtained in the Monte Carlo sampling. Although the functions $K(\tau)$ are negative (well known for the weakly interacting Bose gas~\cite{CapogrossoSansone09}) and lead to a sign problem, 
this only becomes an issue at very low temperatures  in the presence of a condensate (see Tab.~\ref{table}) but does not prevent an accurate computation of phase diagrams and dynamical quantities. 
For the temperatures considered here one iteration takes just a few minutes on a single CPU core.

\begin{figure}[t]
\begin{center}
\includegraphics[angle=0, width=0.5\columnwidth]{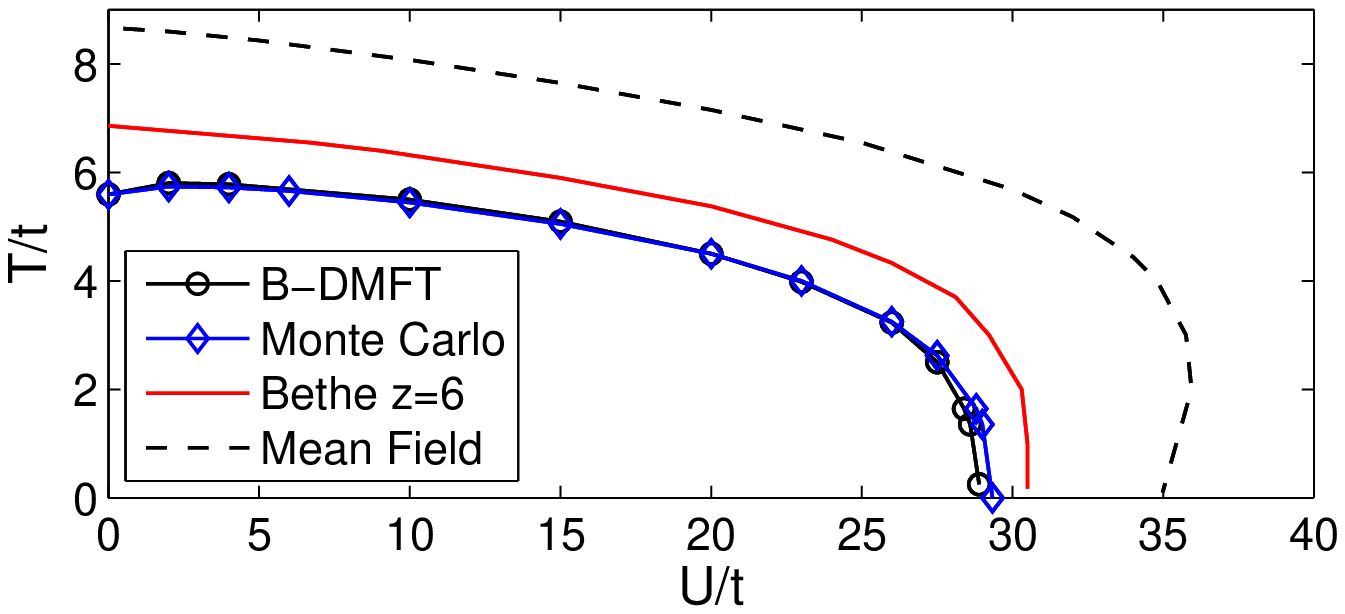}
\includegraphics[angle=0, width=0.5\columnwidth]{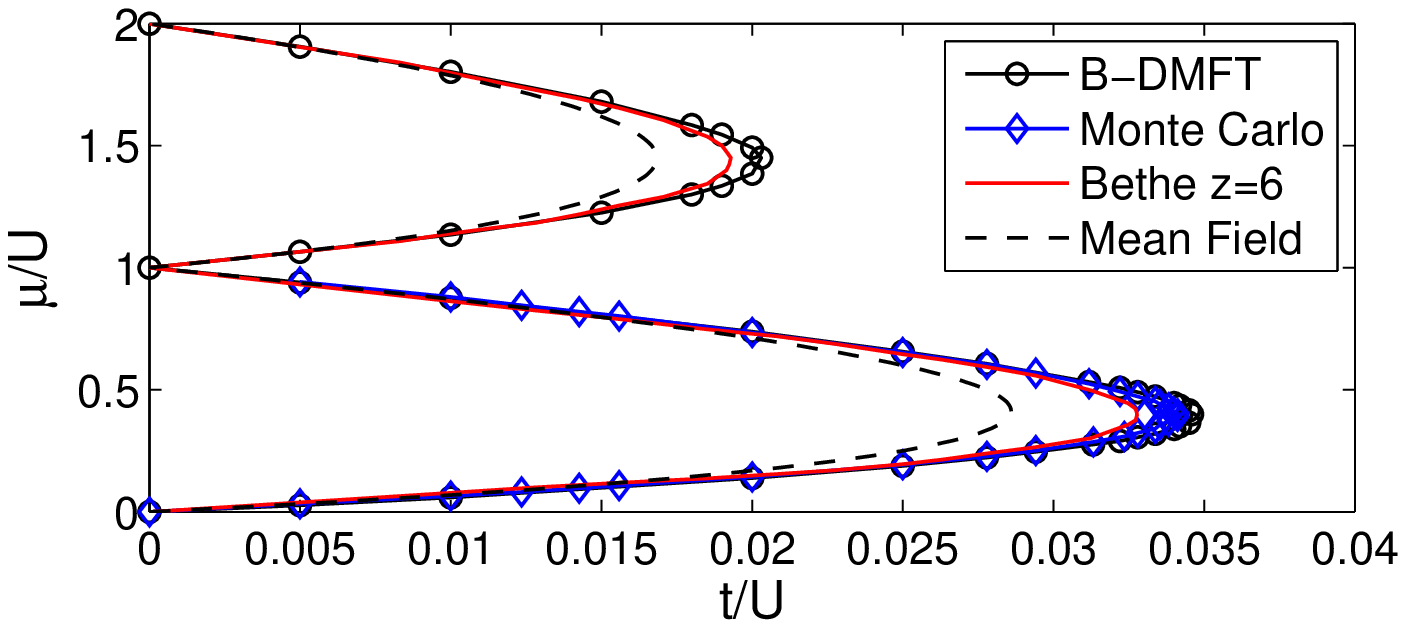}
\caption{Top panel: phase diagram (superfluid to normal liquid transition) in the space of interaction and temperature for $n=1$. The dashed line shows the static mean field result, the red curve the exact solution for a Bethe lattice with coordination number $z=6$ (Ref.~\onlinecite{Semerjian09}) and the blue curve with open diamonds the QMC result from lattice simulations (Ref.~\onlinecite{MC}). The black line with open circles corresponds to the B-DMFT solution, which yields a second order transition.
Bottom panel: ground-state phase diagram in the space of $t/U$ and $\mu/U$, showing the first two Mott lobes surrounded by superfluid. The B-DMFT phase boundary was computed at $\beta t=2$. Error bars are much smaller than the symbol size.}
\label{ground_state}
\end{center}
\end{figure}

As main results we show the finite temperature phase diagram (top panel of Fig.~\ref{ground_state}) and the ground state phase diagram (bottom panel of Fig.~\ref{ground_state}) for the first and second lobe of the Bose Hubbard model on a 3D simple cubic lattice and compare results obtained with B-DMFT to exact results from lattice  QMC simulations, the exact solution for the Bethe lattice with coordination number $z=6$ \cite{Semerjian09}, and to static mean field results. For the calculation of the ground state phase diagram we used $\beta t=2$, which is shown in Fig.~\ref{ground_state} to be a  sufficiently low temperature \footnote{By simulating at $\beta t=4$ and $\beta t=8$ we checked that  any systematic error is smaller than the statistical error.}. 

\begin{figure}[t]
\begin{center}
\includegraphics[angle=0, width=0.25\columnwidth]{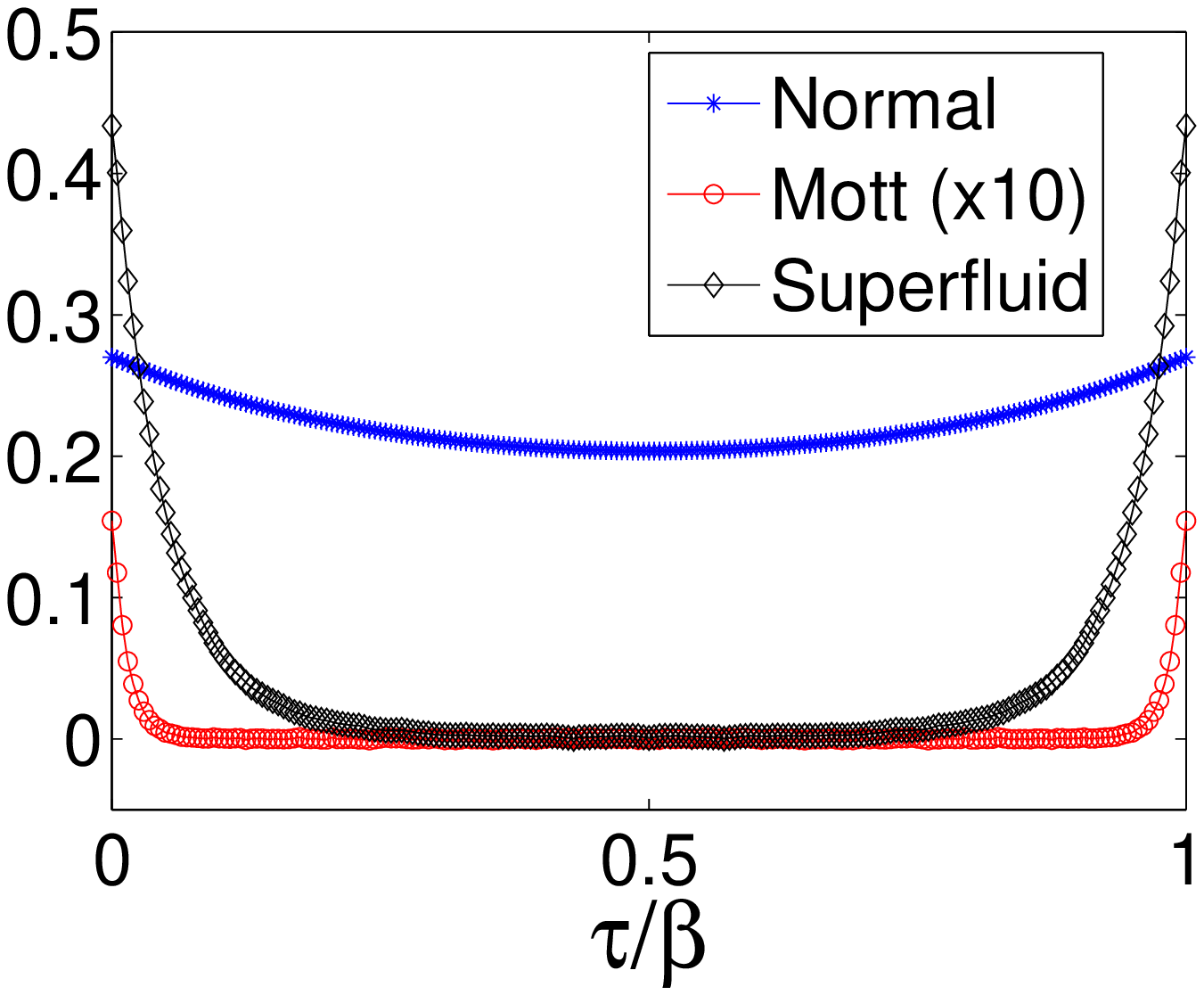}
\includegraphics[angle=0, width=0.25\columnwidth]{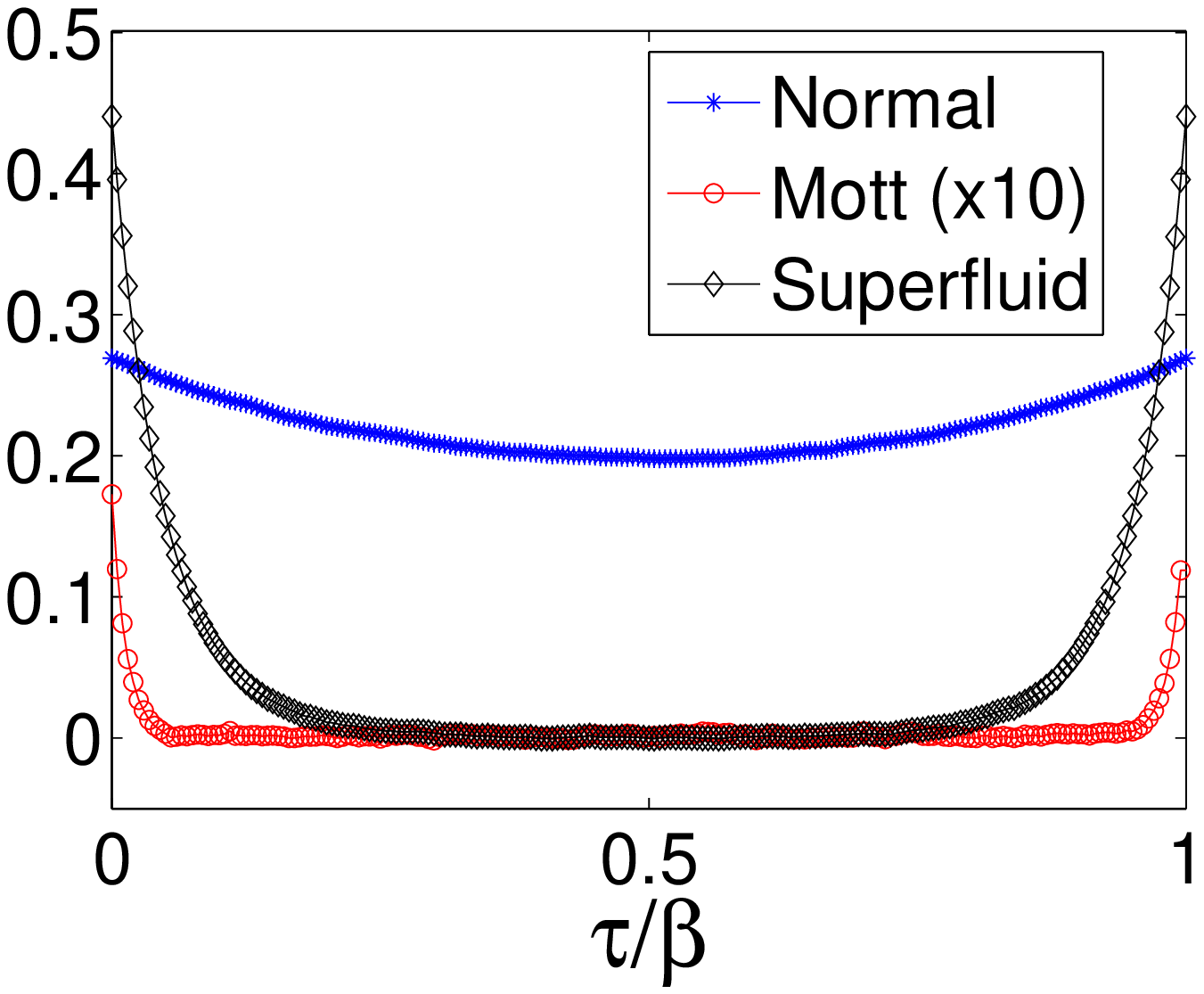}
\caption{Comparison of the connected density-density correlation functions $\langle n(\tau) n(0)\rangle - \langle n \rangle^2$ in the normal ($U$=20, $\mu$=8.72, $\beta$=0.2), superfluid ($U$=10, $\mu$=0.7, $\beta$=2) and Mott insulating ($U$=40, $\mu$=11, $\beta$=2) phases. The left panel shows the B-DMFT result and the right panel the exact data from lattice QMC. The data for the Mott phase are scaled by a factor $10$. 
}
\label{correlators}
\end{center}
\end{figure}

The excellent agreement between our B-DMFT results and the full solution of the Bose-Hubbard model shows that the Mott-transition is a local phenomenon, well described by a momentum-independent self-energy and that the condensed bosons are accurately described by a uniform condensate. B-DMFT also yields remarkably accurate data in the condensed phase, as illustrated in Tab.~\ref{table}. Even dynamical quantities such as density-density correlation functions are correctly reproduced in all three phases (Fig.~\ref{correlators}).

\begin{table}[b]
\centering
\begin{footnotesize}
\begin{tabular}{|l|l|l|l|l|l|}
 \hline
 ($U$, $\mu$)	& $n_{\text{B-DMFT}}$ & $n_{\text{MC}}$ & $\phi^2_{\text{B-DMFT}}$ & $\phi^2_{\text{MC}}$  & \text{sign}\\
 \hline
	(20, 6.6) & $0.99441(4)$ & $0.99456(1)$ & $0.5042(3)$ & $0.486(2)$ & 0.6373(1)\\ 
 \hline
	(24, 8.6) & $0.99494(5)$ & $0.995120(1)$ & $0.3383(4)$ & $0.316(1)$ & 0.7836(1)\\ 
 \hline
	(26, 10) & $1.00194(3)$ & $1.001936(1)$ & $0.2389(4)$ & $0.2227(9)$ & 0.8674(1)\\ 
 \hline
	(28, 11.3) & $1.00252(3)$ & $1.002598(4)$ & $0.1087(5)$ & $0.104(1)$ & 0.9585(1)\\ 
 \hline
	(30, 13) & $1.000403(5)$ & $1.00041(4)$ & $0$ & $0$ & 1\\ 
 \hline
	(32, 15) & $1.000333(5)$ & $1.000370(9)$ & $0$ & $0$ & 1\\ 	
 \hline
\end{tabular}
\caption{Comparison of $n$ and $\phi^2$ between lattice QMC and B-DMFT for different values of $\mu$ and $U$ at $\beta t =1$.}
\label{table}
\end{footnotesize}
\end{table}

The generalization of our QMC algorithm to bosonic mixtures is straightforward. It can also be extended to spinful bosons with spin-dependent or  more complicated interactions using a matrix-formulation analogous to the fermionic algorithm of Ref.~\onlinecite{Werner06Kondo}. Similarly, the method can be extended to impurity clusters to solve cluster-generalizations \cite{Maier05} of B-DMFT. These extensions will enable the simulation of superfluids, supersolids, and super-counter-fluids in bosonic mixtures and spinor condensates and insulators in spinful bosonic systems.

Our bosonic impurity solver, based on a hybridization expansion, can easily be combined with its fermionic counterpart \cite{Werner06}, enabling an efficient DMFT simulation of Bose-Fermi mixtures. While powerful numerical methods exist for the simulation of bosonic lattice models~\cite{Prokofev98}, these simulations become prohibitively difficult as soon as fermions are involved. We thus view the solver and formalism described here not only as an efficient tool for the solution of bosonic problems, but also as an important step which opens the door to a systematic investigation of Bose-Fermi mixtures \cite{Byczuk09}. 

{\it Acknowledgments}
Calculations have been performed on the Brutus cluster at ETH Zurich. 
We acknowledge very helpful discussions with K. Byczuk, A. Georges, E. Kozik, C. May, O. Parcollet, N. Prokof'ev, B. Svistunov, D. Vollhardt, and F. Zamponi, and thank W.-J. Hu, N.-H. Tong, A. Hubener and W. Hofstetter for sharing their B-DMFT results. This project was supported by the Swiss National Science Foundation, NSF under Grant No. DMR-0705847, and by a grant from the Army Research Office with funding from the DARPA OLE program. We also acknowledge hospitality of KITP Santa Barbara and the Aspen Center for Physics.

\newpage
{\Large{Auxiliary material for Dynamical mean field solution of the Bose-Hubbard model}\\}

\subsection{Scope}
In this supplementary material we provide a formal basis for the action and the self-consistency relations that were presented in the manuscript. We implement an expansion around the atomic limit, following almost literally  the lecture notes by A. Georges~\cite{Antoine}, and consider B-DMFT as an approximation to the kinetic energy functional. The atomic reference system is interpreted as our impurity problem. Use is made of the coupling constant integration method and source fields (Lagrange multipliers) are introduced to constrain the condensate field and the connected Green's function for the normal bosons to their physical values. 

\subsection{Expansion parameter}
We introduce a parameter $\alpha \in [0,1]$ such that ($\langle i,j \rangle$ denotes nearest neighbor sites)
\begin{equation}
H_{\alpha} = \frac{U}{2} \sum_i n_i (n_i - 1) - \alpha t \sum_{\langle i,j \rangle} b_i^{\dagger} b_j.
\label{eq:antoine_alpha}
\end{equation}
When $\alpha$ = 0, the atomic limit is recovered and the partition function factorizes over all sites. When $\alpha=1$ the full hopping is recovered, and this is the model we are ultimately interested in.
\subsection{Source fields and constraining fields}

Constraining the normal/anomalous Green's functions and the condensate to specified values can be done by introducing conjugate source fields (Lagrange multipliers) in the action. In order to constrain the condensate to $\mathbf{\Phi}$ we introduce the source field $\mathbf{J}$, and analogous for the {\it connected} Green's function $\mathbf{G}_c$ with source field $\mathbf{\Delta}$. Throughout this document we use the Nambu notation in which $\mathbf{\Phi}^\dagger=(\phi^*, \phi)$, $\mathbf{J}^\dagger=(J^*, J)$ and the individual components of $\mathbf{G}_c$ and $\mathbf{\Delta}$ are given by
\begin{eqnarray}
\mathbf{G}_c(\tau) =  \left( \begin{array}{cc}
G_c(\tau)  & \tilde{G}_c(\tau) \\
\tilde{G}_c^{*}(\tau) &  G_c(-\tau)
\end{array} \right),
\end{eqnarray}
and
\begin{eqnarray}
\mathbf{\Delta}(\tau)= \left( \begin{array}{cc}
F(-\tau)  & 2K(\tau) \\
2K^{*}(\tau) & F(\tau)
\end{array} \right).
\end{eqnarray}
We can then explicitly write down the grand potential per site (there are $N_s$ sites) which is a functional of the source fields and also depends on the constraining fields, 
\begin{eqnarray}
\Omega_{\alpha}[\mathbf{J}, \mathbf{\Phi}, \mathbf{\Delta}, \mathbf{G_c}] & = & -\frac{1}{N_s \beta} \ln \int \mathcal{D}[b^*, b] \exp \Bigg\{ \int_0^{\beta} d\tau \left( \sum_i b_i^* ( -\partial_{\tau} + \mu ) b_i - H_{\rm \alpha}[b^*, b] \right) \nonumber \\
{} & {} & + \int_0^{\beta} d\tau \sum_i \left( J^*(\tau) [ b_i(\tau) - \phi_i(\tau) ] + J(\tau) [b_i^*(\tau) - \phi_i^*(\tau) ] \right)\nonumber \\
{} & {} & + \int_0^{\beta} d\tau \int_0^{\beta} d\tau' \sum_i F(\tau - \tau') [ \delta b_i(\tau) \delta  b_i^*(\tau') + G_c(\tau - \tau') ] \nonumber \\
{} & {} & + \int_0^{\beta} d\tau \int_0^{\beta} d\tau' \sum_i K(\tau - \tau') [ \delta b_i^*(\tau) \delta  b_i^*(\tau') + \tilde{G}_c^*(\tau - \tau') ] \nonumber \\
{} & {} & + \int_0^{\beta} d\tau \int_0^{\beta} d\tau' \sum_i K^*(\tau - \tau') [ \delta b_i(\tau) \delta  b_i(\tau') + \tilde{G}_c(\tau - \tau') ] \Bigg\}.
\end{eqnarray}
\subsection{Atomic limit : impurity model}
Let us consider the case $\alpha = 0$, which is the atomic limit. The problem becomes local on every site with grand potential
\begin{eqnarray}
\Omega_{0}[\mathbf{J}, \mathbf{\Phi}, \mathbf{\Delta}, \mathbf{G_c}] & = & -\frac{1}{N_s \beta} \ln \int \mathcal{D}[b^*, b] \exp \Bigg\{ \int_0^{\beta} d\tau \left( \sum_i b_i^* ( -\partial_{\tau} + \mu ) b_i - \frac{U}{2} n_i(n_i-1) \right) \nonumber \\
{} & {} & + \int_0^{\beta} d\tau \sum_i \left( J_0^*(\tau) [ b_i(\tau) - \phi_i(\tau) ] + J_0(\tau) [b_i^*(\tau) - \phi_i^*(\tau) ] \right)\nonumber \\
{} & {} & + \int_0^{\beta} d\tau \int_0^{\beta} d\tau' \sum_i F_0(\tau - \tau') [ \delta b_i(\tau) \delta  b_i^*(\tau') + G_c(\tau - \tau') ] \nonumber \\
{} & {} & + \int_0^{\beta} d\tau \int_0^{\beta} d\tau' \sum_i K_0(\tau - \tau') [ \delta b_i^*(\tau) \delta  b_i^*(\tau') + \tilde{G}_c^*(\tau - \tau') ] \nonumber \\
{} & {} & + \int_0^{\beta} d\tau \int_0^{\beta} d\tau' \sum_i K_0^*(\tau - \tau') [ \delta b_i(\tau) \delta  b_i(\tau') + \tilde{G}_c(\tau - \tau') ] \Bigg\}.
\end{eqnarray}
From $\delta\Omega_0/\delta J_0=0$ and $\delta\Omega_0/\delta J_0^*=0$ we obtain 
\begin{equation}
\mathbf{\Phi}=\langle\mathbf{b}\rangle_{S_{imp}},
\label{phi}
\end{equation}
and from $\delta\Omega_0/\delta F=0$, $\delta\Omega_0/\delta K=0$, $\delta\Omega_0/\delta K^*=0$ the relation 
\begin{equation}
\mathbf{G}_c(\tau)=\mathbf{G}(\tau)+\mathbf{\Phi}\mathbf{\Phi}^\dagger,
\label{Gc}
\end{equation}
with $\mathbf{G}(\tau)=-\langle T\mathbf{b}(\tau)\mathbf{b}^{\dagger}(0)\rangle_{S_\text{imp}}$. The expectation values $\langle \ldots \rangle_{S_\text{imp}}=Tr[T e^{-S_\text{imp}} \ldots ]/Tr[T e^{-S_\text{imp}}]$ are defined with respect to an impurity action
\begin{eqnarray}
S_{\rm imp} & = & - \frac{1}{2}\int_0^{\beta} \int_0^{\beta} d\tau  d\tau' \delta \mathbf{b}^*(\tau) {\mathbf{\Delta}}_0 (\tau - \tau')  \delta\mathbf{b}(\tau') - \int_0^{\beta} d\tau \mathbf{J}_0^{\dagger}(\tau) \mathbf{b}(\tau) \nonumber\\
&&  
-\mu\int_0^\beta d\tau n(\tau) + \frac{U}{2} \int_0^{\beta} d\tau n(\tau) [ n(\tau) - 1].
\label{imp}
\end{eqnarray}
Inverting expressions (\ref{phi}) and (\ref{Gc}) yields $\mathbf{J}[\mathbf{\Phi}, \mathbf{G_c} ]$ and $ \mathbf{\Delta}_0  [ \mathbf{\Phi}, \mathbf{G_c} ]$ and thus a functional $\Gamma_0$ of the condensate and connected impurity Green's function:
\begin{equation}
\Gamma_0[ \mathbf{\Phi}, \mathbf{G_c} ] = F_\text{imp}[ \mathbf{\Phi}, \mathbf{G_c} ] - \int_0^{\beta} d\tau [F_0(\tau)G_c(\tau)+K_0(\tau)\tilde G_c^*(
\tau)+K_0^*(\tau)\tilde G_c(\tau)]
+\frac{1}{N_s\beta}\int_0^\beta d\tau \sum_i [J_0^*(\tau)\phi_i(\tau)+J_0(\tau)\phi_i^*(\tau)].
\end{equation}
\subsection{Full model}

The exact functional of the (local) Green's function and condensate are constructed using the coupling constant integration method, starting from the atomic limit, $\Gamma = \Gamma_{\alpha=1} = \Gamma_0 + \int_0^1 d\alpha \frac{d\Gamma_{\alpha}}{d\alpha}$.
Using the stationarity of $\Omega$ ($\alpha$-derivatives of the Lagrange multipliers do not contribute),
\begin{eqnarray}
\frac{d\Gamma_\alpha}{d\alpha} & = & -\frac{1}{N_s \beta}  \int_0^\beta d\tau t \sum_{\langle i,j \rangle}  \langle  b_i^*(\tau) b_j(\tau) \rangle =
-\frac{1}{N_s \beta}  \int_0^\beta d\tau t \sum_{\langle i,j \rangle}  [ \phi_i^*(\tau) \phi_j(\tau) 
+ \langle \delta b_i^*(\tau) \delta b_j(\tau)\rangle  ]
 \\
{} & = & \frac{1}{N_s \beta} \text{Tr} \sum_{n, \bf k} \epsilon_{\bf k} \mathbf{G}^{\alpha}_c(\mathbf{k}, i \omega_n) \vert_{ \mathbf{\Phi, G}_c}
-\frac{1}{N_s \beta}  \int_0^\beta d\tau t \sum_{\langle i,j \rangle}  [ \phi_i^*(\tau) \phi_j(\tau) ]. \label{dGda}
\end{eqnarray}
We arrive at the formal expression for the exact functional $\Gamma = \Gamma_{\alpha = 1}$,
\begin{equation}
\Gamma[ \mathbf{\Phi},  \mathbf{G}_c] = \Gamma_0[ \mathbf{\Phi},  \mathbf{G}_c] + \mathcal{K}[ \mathbf{\Phi},  \mathbf{G}_c],
\end{equation}
with the kinetic energy functional $\mathcal{K}[ \mathbf{\Phi},  \mathbf{G}_c]=\int_0^1 d\alpha \frac{d\Gamma_{\alpha}}{d\alpha}[ \mathbf{\Phi},  \mathbf{G}_c]$. 
Requiring stationarity ($\delta \Gamma / \delta \phi_i^*(\tau)=0 $, $\delta \Gamma / \delta \phi_j(\tau)=0 $) determines the value of the source field conjugate to the condensate (assume a homogeneous condensate over the lattice), 
\begin{equation}
\mathbf{J}_0 = zt \mathbf{\Phi}.
\end{equation}
Since the condensate is time-independent (and taken real), we drop the $\tau$ dependence of $J_0$ as well.
The other  stationarity requirement
($\delta \Gamma / \delta G_c=0$, $\delta \Gamma / \delta \tilde G_c=0$) 
determines the hybridization function: 
\begin{equation}
F_0(\tau) = \frac{\delta {\mathcal{K}} }{\delta G_c (\tau) }, \hspace{2mm} K_0(\tau) = \frac{\delta {\mathcal{K}} }{\delta \tilde G_c (\tau) }.
\end{equation}
Note that for the case $z = \infty$, we have identically $\delta b = 0$ and only static mean-field theory exists~\cite{Fisher89}.
\subsection{Approximation to the kinetic energy functional}

B-DMFT can now be considered as an approximation to the kinetic energy functional.
With the single-particle Green's function of the Bose-Hubbard model in the presence of source fields and for arbitrary coupling constants, we can define a selfenergy
\begin{equation}
\mathbf{G}_c^{\alpha}(\mathbf{k}, i\omega_n) = [ i \omega_n\sigma_3 + (\mu - \alpha \epsilon_{\bf k})\mathbf{I}  + \mathbf{\Delta}_{\alpha} [i \omega_n]  - \mathbf{\Sigma}_{\alpha}[{\bf k}, i \omega_n] ]^{-1}.  
\end{equation}
The DMFT approximation consists in replacing the self-energy $\mathbf{\Sigma}_{\alpha}$ for arbitrary $\alpha$ by the impurity model self-energy $\mathbf{\Sigma}_0$. Hence,
\begin{eqnarray}
\mathbf{G}_c^{\alpha}(\mathbf{k}, i\omega_n) \vert_{\text {B-DMFT}} &=& [ i \omega_n\sigma_3 + (\mu- \alpha \epsilon_{\bf k})\mathbf{I}  + \mathbf{\Delta}_{\alpha} [i \omega_n; \mathbf{\Phi}, \mathbf{G} _c ] - \mathbf{\Sigma}_{\alpha = 0}[i \omega_n; \mathbf{\Phi}, \mathbf{G} _c  ] ]^{-1} \\
&=& [\mathbf{\Delta}_{\alpha}-\mathbf{\Delta}_{0}+\mathbf{G} _c^{-1}-\alpha\epsilon_{\mathbf{k}} \mathbf{I}]^{-1},
\end{eqnarray}
where we have used that the impurity self-energy satisfies the Dyson equation 
$\mathbf{\Sigma}_{\alpha = 0}[i \omega_n; \mathbf{\Phi}, \mathbf{G} _c  ] =  
i \omega_n\sigma_3 + \mu\mathbf{I} + \mathbf{\Delta}_0[i \omega_n; \mathbf{\Phi}, \mathbf{G} _c ] - \mathbf{G}_c^{-1}.$
Summing over ${\bf k}$, and using the constraint on the local lattice Green's function, we obtain the following relation between $\mathbf{G}_c$ and the hybridization function: 
\begin{eqnarray}
\mathbf{G}_c(i \omega_n) & = & \int d\epsilon D(\epsilon) (\mathbf{\zeta} - \alpha \epsilon\mathbf{I})^{-1} 
= \frac{1}{\alpha} \tilde{D}\Big(\frac{\mathbf{\zeta}}{\alpha} \Big),
\end{eqnarray}
with $\mathbf{\zeta}  =  \mathbf{\Delta}_{\alpha} - \mathbf{\Delta}_{0}  + \mathbf{G}_c^{-1}$.
We used the non-interacting density of states is $D(\epsilon) = \frac{1}{N_s} \sum_{\bf k} \delta(\epsilon - \epsilon_{\bf k})$ and its Hilbert transform $\tilde{D}(\mathbf{z}) = \int d\epsilon D(\epsilon) (\mathbf{z}-\epsilon \mathbf{I})^{-1}$.
By introducing its inverse, $\tilde{D}( R(g)) = g$, the relation above can be inverted ($\alpha R(\alpha \mathbf{G}_c) = \mathbf{\zeta} =  \mathbf{\Delta}_{\alpha} - \mathbf{\Delta}_{0}  + \mathbf{G}_c^{-1}$) and yields the hybridization function as a functional of the local Green's function,
\begin{equation}
\mathbf{\Delta}_{\alpha} [i \omega_n; \mathbf{\Phi}, \mathbf{G}_c] = -\mathbf{G}_c^{-1} +  \mathbf{\Delta}_{0}[\mathbf{\Phi}, \mathbf{G}_c] + \alpha R(\alpha \mathbf{G}_c).
\end{equation}
The lattice Green's function expressed as a functional of $\mathbf{G}_c$ becomes 
\begin{equation}
\mathbf{G}_c^{\alpha}({\bf k}, i\omega_n) \vert_{\text {B-DMFT}}= (\alpha R(\alpha \mathbf{G}_c) - \alpha \epsilon_{\bf k}\mathbf{I})^{-1}.
\end{equation}
Equation (\ref{dGda}) can now be evaluated with $\mathbf{G}_c^\alpha(\mathbf{k}) \vert_{\text {B-DMFT}}$:
\begin{eqnarray}
&&\frac{1}{N_s \beta} \text{Tr} \sum_{n, \bf k} \epsilon_{\bf k} \mathbf{G}_c^{\alpha}({\bf k}, i\omega_n) \vert_{\text {B-DMFT}} -\frac{1}{N_s \beta}  \int_0^\beta d\tau t \sum_{\langle i,j \rangle}  [ \phi_i^*(\tau) \phi_j(\tau) ] \nonumber\\
& &\hspace{10mm}= \frac{1}{\alpha} \frac{1}{\beta}\sum_n \text{Tr} \int d\epsilon \epsilon D(\epsilon) (R(\alpha \mathbf{G}_c )- \epsilon\mathbf{I})^{-1} -\frac{1}{N_s \beta}  \int_0^\beta d\tau t \sum_{\langle i,j \rangle}  [ \phi_i^*(\tau) \phi_j(\tau) ] \nonumber. \nonumber \\
&&\hspace{10mm}= \frac{1}{\alpha} \frac{1}{\beta}\sum_n \text{Tr} \Big[ -\mathbf{I} + \alpha \mathbf{G}_c R(\alpha \mathbf{G}_c ) \Big] -\frac{1}{N_s \beta}  \int_0^\beta d\tau t \sum_{\langle i,j \rangle}  [ \phi_i^*(\tau) \phi_j(\tau) ].
\end{eqnarray}
An explicit expression for the B-DMFT approximation to $\mathcal{K}[\mathbf{\Phi}, \mathbf{G}_c]$ therefore reads
\begin{equation}
\mathcal{K}_{\text{B-DMFT}}[\mathbf{\Phi}, \mathbf{G}_c] =  \int_0^1 d\alpha \frac{1}{\beta}\sum_n \text{Tr}\left[   \mathbf{G}_c(i \omega_n) R(\alpha \mathbf{G}_c(i \omega_n) ) - \alpha^{-1}\mathbf{I}  \right] -\frac{1}{N_s \beta}  \int_0^\beta d\tau t \sum_{\langle i,j \rangle}  [ \phi_i^*(\tau) \phi_j(\tau) ],
\label{ekinbdmft}
\end{equation}
where the last term reduces to $-zt \phi^* \phi$ for a constant, homogeneous condensate.
 
\subsection{Stationarity conditions}

It immediately follows from Eq.~(\ref{ekinbdmft}) that the stationarity condition for the condensate is unaltered in the B-DMFT approximation ($\mathbf{J}_0 = zt \mathbf{\Phi}$), while the stationarity condition for the connected Green's function ($\delta \Gamma / \delta G_c=0$, $\delta \Gamma / \delta \tilde G_c=0$) reads in the B-DMFT approximation (use $R(\alpha G) + \alpha GR'(\alpha G) = \partial_{\rm \alpha}[\alpha R(\alpha G)]$ and the cyclical properties of the trace),
\begin{equation}
\mathbf{\Delta}_0[i \omega_n; \mathbf{\Phi}, \mathbf{G}_c] \vert_{\text{B-DMFT}} = -R[ \mathbf{G}_c(i \omega_n) ] + \mathbf{G}_c (i \omega_n)^{-1} = -i\omega_n\sigma_3-\mu\mathbf{I}+\mathbf{\Sigma}_{imp}+\mathbf{G}_c^{-1}.
\label{hyb}
\end{equation}
Applying $\tilde{D}(.)$ to both sides of Eq.~(\ref{hyb}) gives
\begin{equation}
\mathbf{G}_c ( i \omega_n) = \int d\epsilon D(\epsilon)( i \omega_n\sigma_3 + (\mu-\epsilon)\mathbf{I} - \mathbf{\Sigma}_{\rm imp}( i \omega_n) )^{-1}.
\label{selfconsistency}
\end{equation}
This equation defines the B-DMFT self-consistency condition.
\subsection{Summary}

B-DMFT maps the bosonic lattice problem to a self-consistent solution of an impurity model, whose action (expressed in terms of the full operators $\bf{b}$) reads

\begin{eqnarray}
S_{\rm B-DMFT} & = & - \frac{1}{2}\int_0^{\beta} \int_0^{\beta} d\tau  d\tau' \mathbf{b}^*(\tau) {\mathbf{\Delta}}_0 (\tau - \tau') \mathbf{b}(\tau') -\mu\int_0^\beta d\tau n(\tau) + \frac{U}{2} \int_0^{\beta} d\tau n(\tau) [ n(\tau) - 1]  \nonumber\\
&&  - \mathbf{\Phi}^{\dagger} \Big(zt-\int_0^\beta d\tau'  {\mathbf{\Delta}}_0 (\tau') \Big) \int_0^{\beta} d\tau \mathbf{b}(\tau).
\label{imp}
\end{eqnarray}
Using the fact that $K=K^*$ can be chosen real, one recovers the action presented in the manuscript.
The solution of the impurity problem yields the condensate $\mathbf{\Phi}$ (Eq.~(\ref{phi})), the connected Green's function $\mathbf{G}_c$ (Eq.~(\ref{Gc})) and the self-energy $\mathbf{\Sigma}_{imp}$ of the impurity model. The right hand side of Eq.~(\ref{selfconsistency}) then defines the local lattice Green's function, which is identified with the impurity Green's function and thus allows to define the new hybridization function for the next iteration using Eq.~(\ref{hyb}).


\begin{thebibliography}{99}
\bibitem{Metzner89} W. Metzner and D. Vollhardt, Phys. Rev. Lett. {\bf 62}, 324 (1989).
\bibitem{Georges92} A. Georges and G. Kotliar, Phys. Rev. B {\bf 45}, 6479 (1992).
\bibitem{Georges96} A. Georges {\it et al.}, Rev. Mod. Phys. {\bf 68}, 13 (1996).
\bibitem{Kotliar06} G. Kotliar {\it et al.}, Rev. Mod. Phys. {\bf 78}, 865, (2006).
\bibitem{Chitra01} R. Chitra and G. Kotliar, PRB {\bf 63}, 115110 (2001).
\bibitem{Fisher89} M. P. A. Fisher, P. B. Weichman, G. Grinstein, and D. S. Fisher, Phys. Rev. B {\bf 40}, 546 (1989).
\bibitem{Byczuk08} K. Byczuk and D. Vollhardt, Phys. Rev. B \textbf{77}, 235106 (2008).
\bibitem{Hu09} W.-J. Hu and N.-H. Tong, Phys. Rev. B {\bf 80}, 245110 (2009).
\bibitem{Hubener09} A. Hubener, M. Snoek, W. Hofstetter, Phys. Rev. B {\bf 80}, 245109 (2009).
\bibitem{CapogrossoSansone09} B. Capogrosso-Sansone {\it et al.},  New J. Phys. {\bf 12} 043010 (2010).
\bibitem{Prokofev98} N. V. Prokof'ev, B. V. Svistunov, and I. S. Tupitsyn, JETP {\bf 87}, 310 (1998); L. Pollet, K. Van Houcke, and S. Rombouts, J. Comp. Phys. {\bf 225}, 2249 (2007).
\bibitem{MC} B. Capogrosso-Sansone, N.V. Prokof'ev and B.V. Svistunov, Phys. Rev. B {\bf 75}, 134302 (2007).
\bibitem{Antoine} A. Georges, in ``LECTURES ON THE PHYSICS OF HIGHLY CORRELATED ELECTRON SYSTEMS VIII:  Eighth Training Course in the Physics of Correlated Electron Systems and High-Tc Superconductors" Vol. {\bf 715}, p. 3-74, Eds. F. Mancini and A. Avella, AIP (2004). 
\bibitem{EPAPS} See supplementary material at \url{http://link.aps.org/supplemental/10.1103/PhysRevLett.105.096402} for a formal derivation based on an approximation of the kinetic energy functional; or see the Appendix in the arXiv version. 
\bibitem{Rubtsov05} A. N. Rubtsov, V. V. Savkin and A. I. Lichtenstein, Phys. Rev. B {\bf 72}, 035122 (2005).
\bibitem{Gull08} E. Gull {\it et al.},  Europhys. Lett. {\bf 82} 57003 (2008).
\bibitem{Werner06} P.~Werner {\it et al.}, Phys. Rev. Lett. {\bf 97}, 076405 (2006).
 \bibitem{footnote} This approach is different from the Baym-Kadanoff formalism, which is based on a functional of all components of the lattice Green function.
\bibitem{NegeleOrland} J. W. Negele and H. Orland, Quantum Many-Particle Systems, Westview Press, Boulder, 1998. 
\bibitem{Semerjian09} G. Semerjian, M. Tarzia and F. Zamponi, Phys. Rev. B {\bf 80}, 014524 (2009).
\bibitem{Semerjian10}G. Biroli, G. Semerjian and M. Tarzia, arXiv:1005.0342.
\bibitem{Werner06Kondo} P.~Werner and A.~J.~Millis, Phys. Rev. B {\bf 74}, 155107 (2006). 
\bibitem{Maier05} T. Maier {\it et al.}, Rev. Mod. Phys. {\bf 77}, 1027 (2005).
\bibitem{Byczuk09} K. Byczuk and D. Vollhardt, Ann. Phys. (Berlin) {\bf 18}, 622 (2009).



\end{thebibliography}
\end{document}